\documentclass{iau-JDSS}
\usepackage{graphicx}

\title[Stars vs. hot gas in Ellipticals] %% give here short title %%
{Abundance Ratios in Stars vs. Hot Gas in Elliptical Galaxies %: the Chemical Evolution Modeller Point of View
}

\author[A.Pipino]   %% give here short author list %%
{Antonio Pipino$^{1}$}
\affiliation{$^1$ Universita'di Trieste, Italy \& UCLA, USA}

\pubyear{2009}
\volume{Volume 15}  %% insert here IAU Highlights of Astronomy Volume Number
\pagerange{119--126}
\date{?? and in revised form ??}
\jname{Highlights of Astronomy, Volume 15}
\editors{Ian F Corbett, ed.}
\begin{document}

\maketitle

\begin{abstract}
I present predictions from a chemical evolution model aimed at a self-consistent
study of both optical (i.e. stellar) and X-ray (i.e.gas) properties
of present-day elliptical galaxies. Detailed cooling and heating processes in
the interstellar medium (ISM) are taken into and allow a reliable modelling
of the SN-driven galactic wind. The model simultaneously reproduces the mass metallicity,
the colour-magnitude, the L$_x$ - L$_B$ and the L$_X$ - T relations, as
well as the observed trend of the [Mg/Fe] ratio as a function of $\sigma$. The "iron discrepancy" can be solved by taking into account the existence of dust.
%% add here a maximum of 10 keywords, to be taken form the file <Keywords.txt>
\end{abstract}
\firstsection
\begin{figure}[h]
\includegraphics[width=6cm,height=3.5cm]{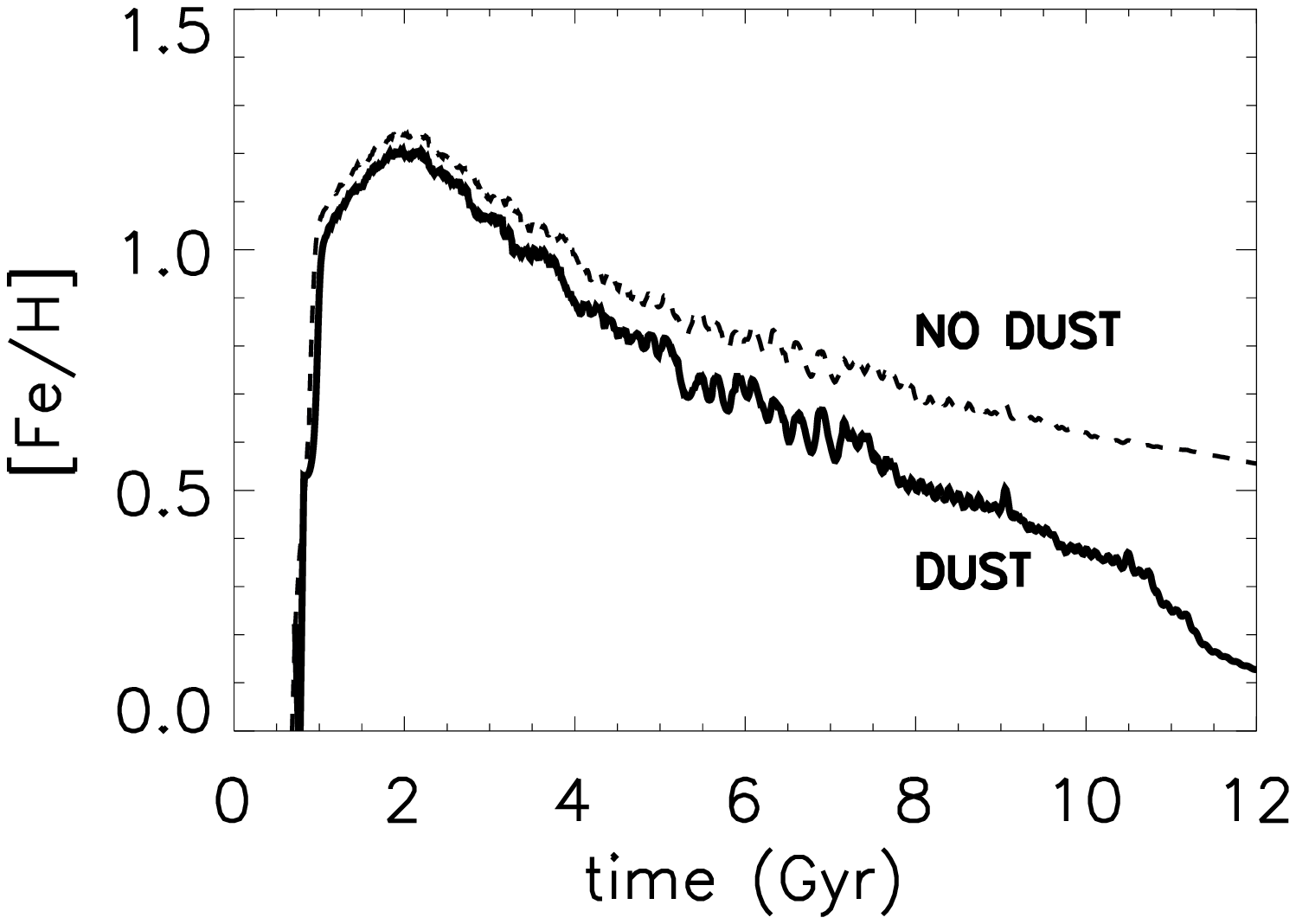}
\includegraphics[width=6cm,height=3.5cm]{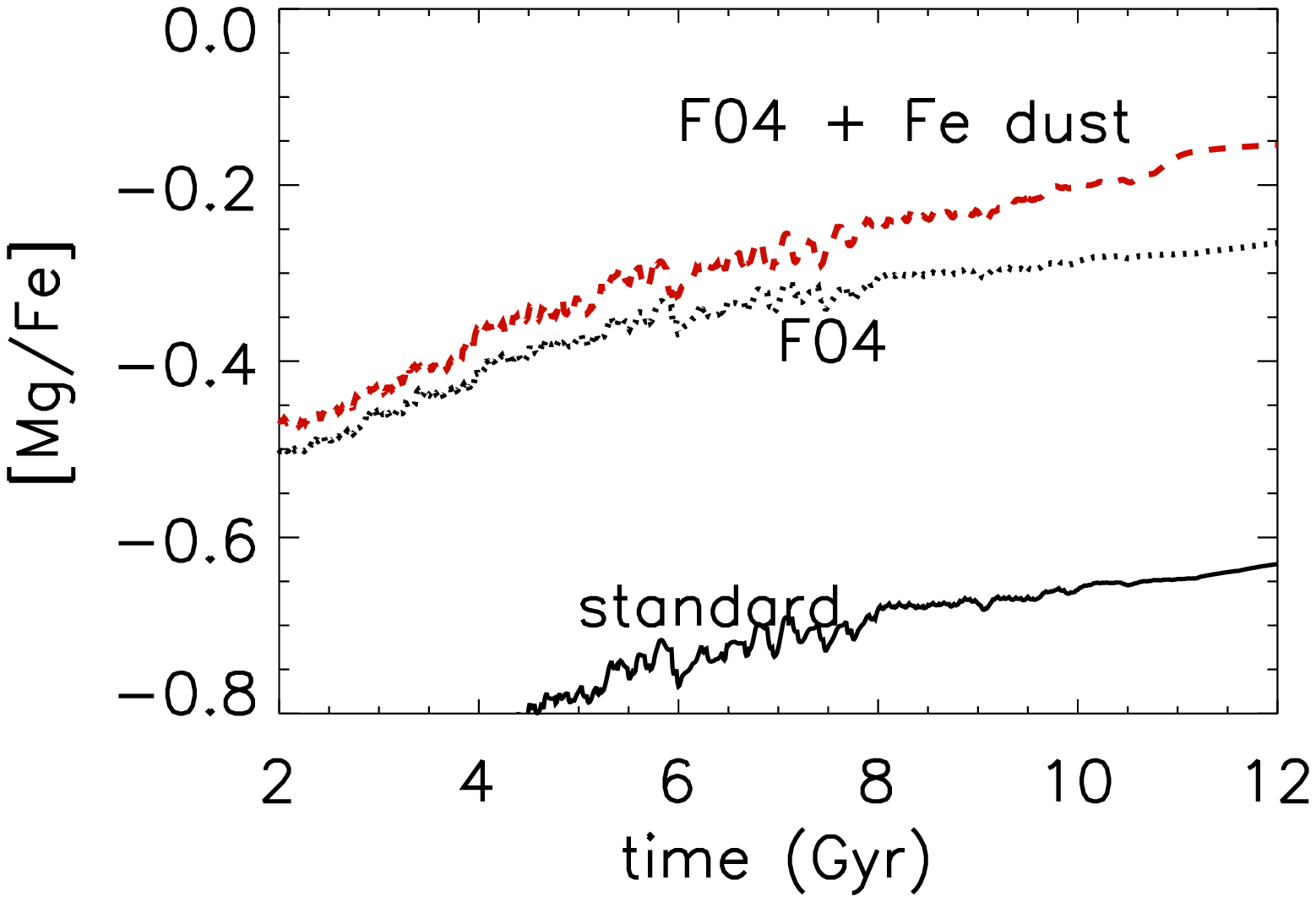}
\caption{Predicted abundance ratios as a function of time by different models (see text).}
\label{fig5}
\end{figure}
Monolithic collapse models featuring a SN-driven galactic wind (Pipino et al., 2008) are shown to reproduce the largest
number of observables in the optical spectrum of elliptical galaxies (e.g. Pipino et al., 2005, P05).
Here, I made use of the P05 chemical evolution code in order to
present preliminary attempts to overcome long lasting problems such as the discrepancy
between the expected high Fe abundance in the post-wind phase and the observed one,
as well as to explain the observed abundance ratio pattern (see Bregman, Humphrey, this Conference).

In particular, in Calura et al. (2008) we showed that the most
recent estimates of the diffuse dust in ellipticals is enough to hide a suitable
amount of Fe and reduce the gas phase abundance to the required solar value (Fig.~\ref{fig5}, left panel).
The empirical yields by Fran\c cois et al. (2004, F04), instead, make the predicted [Mg/Fe]
closer to the observed solar value (Fig.~\ref{fig5}, right panel).
The same yields explain why the observations hint toward an under-solar [O/Mg] ratio.

%\begin{acknowledgments}
I acknowledge support from the AAS and the IAU travel grants.
%\end{acknowledgments}

\end{document}